\documentclass[acus]{JAC2003}

\usepackage{graphicx}
\usepackage{booktabs}
\usepackage{hyperref}

\begin{document}
\title{ POLARIZED POSITRON SOURCE WITH A COMPTON MULTIPLE INTERACTION POINT LINE}

\author{I. Chaikovska\thanks{chaikovs@lal.in2p3.fr}, R. Chehab, O. Dadoun, P. Lepercq, A. Variola\\ Laboratoire de l'Acc\'el\'erateur Lin\'eaire (LAL), Universit\'e Paris-Sud XI, Orsay, France\\
}

\maketitle
\begin{abstract}

Positron sources are critical components of the future linear collider projects. This is essentially due to the high luminosity required, orders of magnitude higher than existing ones. In addition, polarization of the positron beam rather expands the physics research potential of the machine. 
In this framework, the Compton sources for polarized positron production are taken into account where the high energy gamma rays are produced by the Compton scattering and subsequently converted into the polarized electron-positron pairs in a target-converter. The Compton multiple Interaction Point (IP) line is proposed as one of the solutions to increase the number of the positrons produced. 
The gamma ray production with the Compton multiple IP line is simulated and used for polarized positron generation. Later, a capture section based on an adiabatic matching device (AMD) followed by a pre-injector linac is simulated to capture and accelerate the positron beam. 

\end{abstract}
\section{Introduction}
\begin{figure}[htb]
   \centering
   \includegraphics*[width=80mm]{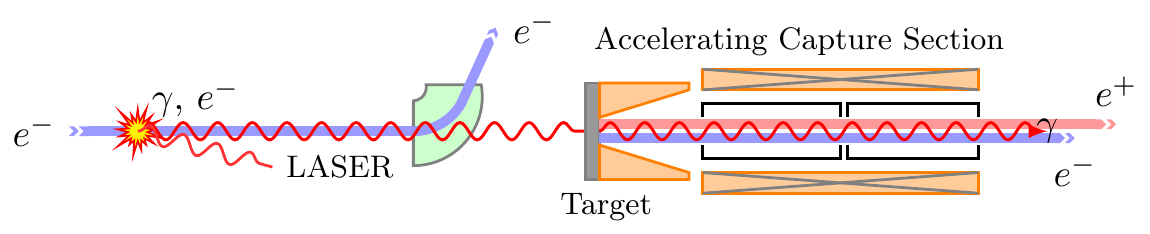}
   \caption{A fundamental scheme of the polarized positron production by the Compton scattering.}
   \label{fig:ComptonScheme}
\end{figure}

Polarized positron source based on the Compton scattering is one of the attractive methods to produce polarized positrons in the case of the future linear collider projects such as the ILC and CLIC~\cite{omori}. A general layout  of this scheme is shown on Fig.~\ref{fig:ComptonScheme}. Polarized laser photons are scattered off the electron beam producing the high energy polarized gamma rays. The later is directed to the target-converter resulting in the production of the polarized positrons which are collected and accelerated by Accelerating Capture Section (ACS) consisting of the matching device to capture the produced positrons and the pre-injector linac for primary acceleration. Then, the positron beam is further accelerated and injected in a Damping Ring (DR).

The main advantages of the Compton scheme are that the positron source is imposed independently with respect to the main linac and the required drive electron beam
energy is much lower as compered with the undulator scheme.
However,  it suffers from the relatively low value of the scattering cross section resulting in the low number of positrons produced per one electron beam crossing. 

Eventually, the challenge of the Compton based Positron Source (CPS) is that, despite the low value of the cross section, to have enough gamma rays to produce enough positrons per second, that is~$\sim\!(1-3)\times10^{14}$~$e^+$~s$^{-1}$ as required by the CLIC and ILC. To meet the requirements imposed by these future projects, one needs to increase significantly the flux of the gamma rays produced. 
This can be done by using a high average power laser system based on a Fabry-Perot cavity and a high current electron beam.  Another solution that should allow an important increase of the emitted gamma ray flux is to design a Compton multiple IP line. This requires a system of focusing triplets to focalise the electron beam on each IP.

\section{Positron production with Compton scattering}
\subsection{Gamma rays production}
There are three types of the CPS according to the electron source used for the Compton scattering: linac based, storage ring based and Energy Recovery Linac (ERL) based CPS. The ERL scheme for CPS is very attractive since the electron bunch after the interaction with the laser pulses is renewed. This allows to have a not degraded bunch for the Compton collisions in every turn which is quite important especially in the case of multiple IP line~\cite{variolaerl}. Therefore, for this study we consider the CPS is composed of the ERL generating the electron beam followed by the ACS. The parameters of the Compton IP used for the simulations are summarized in Table~\ref{table:ComIP}. 

\begin{table}[hbt]
\centering
\small
\caption{Compton IP parameters}
\begin{tabular}{lcc}
\toprule
\textbf{Description} & \textbf{Value} \\ 
Electron energy, $E_e$ & $2$ GeV  \\ 
Electron bunch charge, $C_e$  & $0.5$  nC\\ 
Electron bunch length, $\tau_{e}$  & $3$~ps    \\ 
Electron IP $\beta_x/\beta_y$ & $0.1$ m/rad  \\ 
RMS energy spread, $\delta_E$ & $0.002$  \\
Electron emittance, $\gamma\epsilon_x/\gamma\epsilon_y$ & $10$ $\mu$m rad  \\ 
Electron IP beam size, $\sigma_x/\sigma_y$ & $16 \, \mu m$   \\ 
LASER photon energy, $E_{ph}$ & $1.17$ eV  \\
LASER beam waist, $\omega_L=2\sigma_L$ & $30 \, \mu m$   \\ 
LASER pulse length, $\tau_L$ & $5 $~ps    \\
LASER pulse energy, $E_L$ & $0.6$~J   \\ 
Crossing angle, $\theta$ & $2^{\circ}$   \\ 
Length of one IP, $L_{IP}$ & $5.4$~m   \\ 
 \bottomrule
 \end{tabular}
\label{table:ComIP}
\end{table}

\begin{figure}[htb]
   \centering
   \includegraphics*[width=62mm]{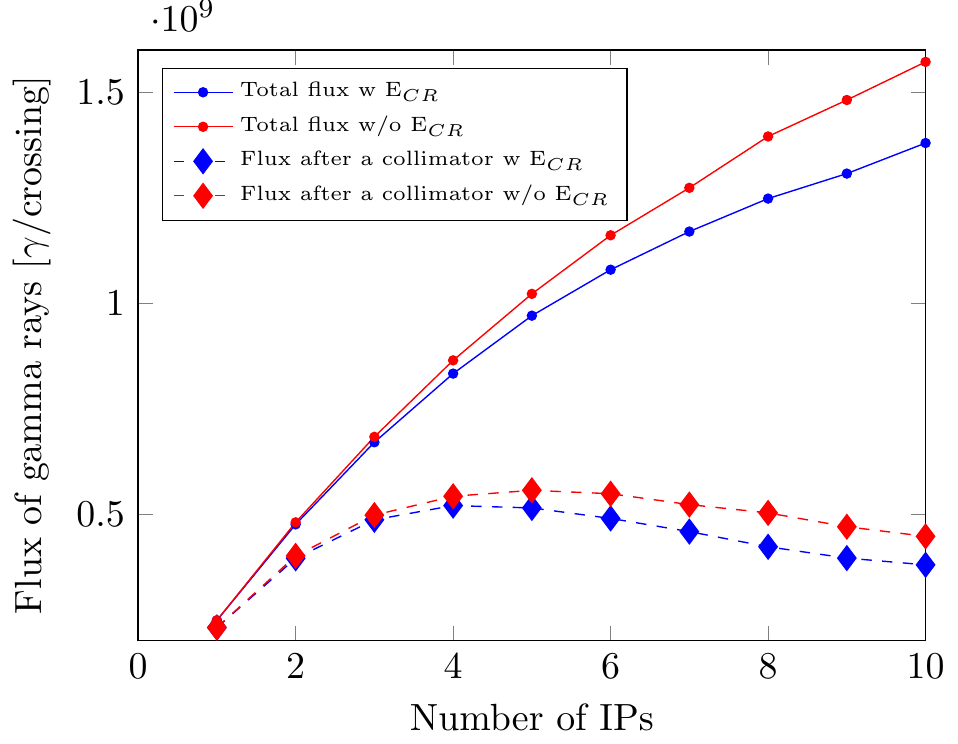}
   \caption{Total gamma ray flux and gamma ray flux obtained after the collimator as a function of the number of the Compton IPs.}
   \label{fig:multiIPline}
\end{figure}

The simulations of the Compton multiple IP line are done inside a code CAIN~\cite{cain} and show a good increasing of the gamma ray flux ( see Fig~\ref{fig:multiIPline}).
 From Fig.~\ref{fig:multiIPline} one can see also that after a few IPs, chromatism of the electron beam caused by the electron energy spectrum degradation
due to the Compton collisions (so-called Compton recoil~$E_{CR}$) start to reduce the gain.
The effect is magnified by the losses due to the collimation system of the gamma rays installed to reach the degree of polarization required.  

It is still efficient to consider 5~IPs where we assumed two lasers cross symmetrically in respect to the beam propagation axis.  In this case a factor four of gain is obtained.
The further optimization of the focusing quadrupole triplets should be done.
\begin{figure}[htb]
   \centering
   \includegraphics*[width=62mm]{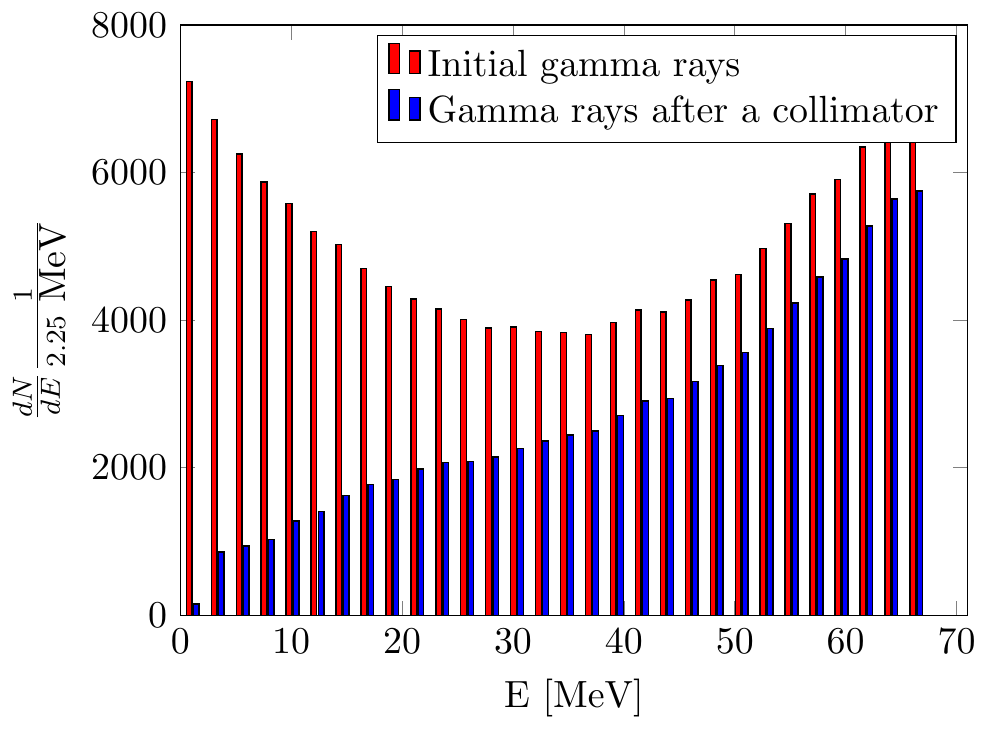}
   \caption{Energy spectrum of the gamma rays produced using the 5~IPs line. . }
   \label{fig:gammaSpectrum}
\end{figure}
The energy spectrum of the gamma rays produced after the 5~IPs is shown on Fig.~\ref{fig:gammaSpectrum} together with the spectrum selected by the collimator to be used for positron generation. The gamma ray mean energy after the collimator is 45~MeV while the gamma ray polarization equals to~37.9\%.

\subsection{Target simulation}
\begin{figure}[htb]
   \centering
   \includegraphics*[width=61mm]{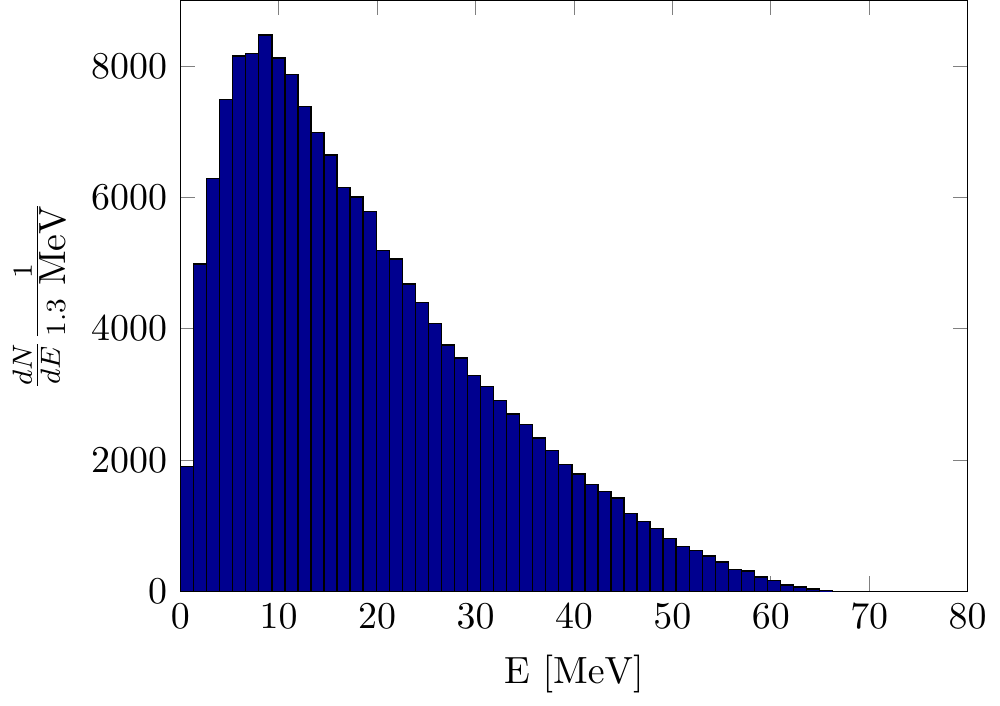}
   \caption{Energy spectrum of the positrons at the exit of the target.}
   \label{fig:posSpectrum}
\end{figure}

Simulations of the positron production is performed using a Geant4 code~\cite{g4} developed at LAL. After the optimization studies regarding the positron yield and polarization, a~4~mm thick target made of Tungsten have been used to produce the polarized positrons. The resulting positron energy spectrum is shown on Fig.~\ref{fig:posSpectrum}. 
The output of this simulation have been used as an input data for the ACS simulation. 


\subsection{Accelerating Capture Section}
The ACS section is presented by the AMD as the matching device and 2~GHz pre-injector linac made of the~17~MV/m constant gradient Travelling Wave (TW) structures to accelerate the positrons up to the $200$~MeV. The simulation of the ACS is performed using a tracking code ASTRA~\cite{astra}.

\begin{figure}[htb]
   \centering
   \includegraphics*[width=60mm]{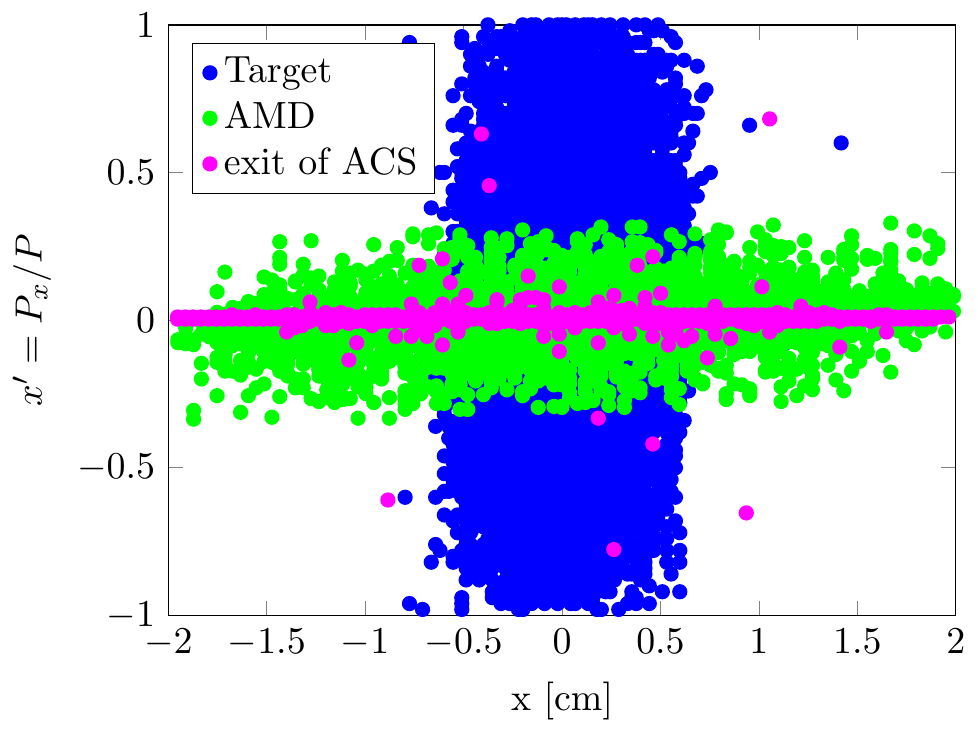}
   \caption{Positron beam emittance taken at the exit of the target, at the exit of the AMD and at the end of the ACS.}
   \label{fig:TarAmdACS}
\end{figure}

The AMD is made of the tapered solenoid field and used to match the positron beam after the production to the acceptance of the pre-injector linac. This can be observed in Fig.~\ref{fig:TarAmdACS}. A positron transverse emittance is transformed in the AMD in such a way that large transverse divergence of the positron beam is decreased and further compressed by the accelerating cavities. 
The AMD used in the simulation is 20~cm long with a longitudinal magnetic field starting at 6~T and adiabatically decreasing down to~0.5~T. The aperture radius of the AMD is 20~mm. The advantage of using the AMD to quarter wave transformer (QWT) as a matching device is that it allows increase the accepted positron yield by capturing the positrons within a wide energy band. 

At the end of the AMD a six 2~GHz TW structures of the pre-injector linac are installed. Each such RF structure is made of 84 cells and 2 couplers with the total length of~4.36~m. The iris radius of the RF cavities is 20~mm. A drift between the structures is set to one $\lambda_{RF}$ that is about~15~cm.
The whole pre-injector linac is encapsulated inside a solenoid with the axial magnetic field equals to~0.5~T. It is employed to avoid the positron losses caused by the high divergence of the positron beam at the beginning of the ACS.

 \begin{figure*}[tb]
    \centering
    \small
   \includegraphics*[height=5.0cm]{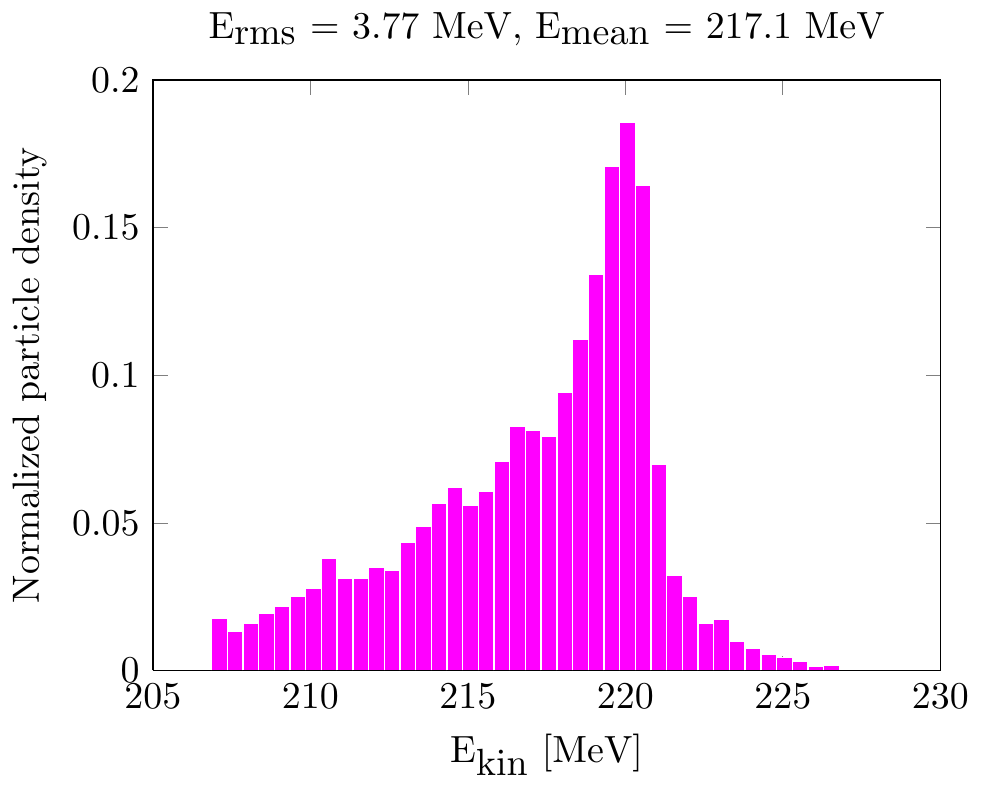} \hspace{0.5cm}
    \includegraphics*[height=5.0cm]{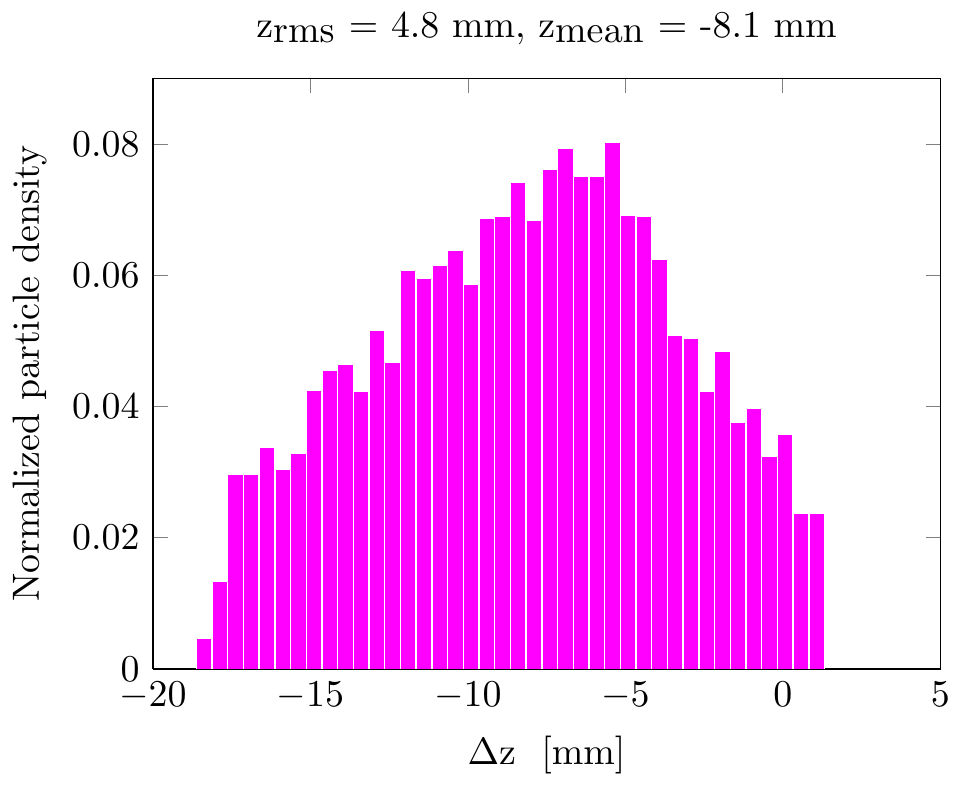}
    \caption{Longitudinal distributions of the positrons at the end of the ACS inside the ($\pm$10~MeV,~$\pm$10~mm) window.}
    \label{fig:posACS}
\end{figure*}

\section{Results and discussion} 
To define the CPS efficiency, a window with ($\pm$10~MeV,~$\pm$10~mm) around the highest density of positrons is set at the end of the ACS. The longitudinal distributions of the positrons selected by this window is shown on Fig.~\ref{fig:posACS}.
The main results of the CPS simulation are presented in Table~\ref{table:results}. According to them, the CPS efficiency normalized by the electron bunch charge and the energy of the laser pulse in the 5~IPs with two crossed lasers scheme is estimated to be~1.71$\times$10$^{-2}$/(nC$\cdot$J) or 5.32$\times$10$^{7}$$e^+$/(nC$\cdot$J). A low positron charge produced per bunch can be compensated by the multiple injections in the same DR RF bucket. In the case of the CLIC requirements~\cite{clic}~$3.72\times10^{9}$~$e^+$/bunch, it is equivalent to 70~injections/(nC$\cdot$J). Assuming the reasonable injection efficiency of 10~injections per DR RF bucket, it is necessary to provide the electron bunches and laser pulses with (nC$\cdot$J) $\approx$~7 to the the Compton IP what is so far not achievable with the current technologies. However, the different R\&D programs are ongoing to improve the performance of the laser systems~\cite{fpc,fpc2} and to ensure the ERL operation in the nC per bunch range.

\begin{table}[hbt]
\centering
\caption{Compton based positron source parameters}
\begin{tabular}{lcc}
\toprule
\textbf{Description} & \textbf{Value} \\ 
Gamma ray production efficiency ($N^{tot}_{\gamma}/N_{e^-}$)  & $0.31$    \\ 
Positrons production yield ($N_{e^+}/N^{ins}_{\gamma}$) & 0.2  \\
Mean positron energy at the production& 18.9 MeV  \\ 
Mean positron polarization at the production & 35.6\%  \\
AMD capture efficiency ($N^{amd}_{e^+}/N^{targ}_{e^+}$) & 0.43  \\
Compton source accepted efficiency($N^{amd}_{e^+}/N_{e^-}$)  & 0.014 \\
ACS efficiency ($N^{\sim200MeV}_{e^+}/N^{targ}_{e^+}$) & 0.24   \\
Compton source efficiency($N^{\sim200MeV}_{e^+}/N_{e^-}$) & 0.0077 \\
ACS efficiency ($N^{\pm10 MeV \pm10mm}_{e^+ (\pm10 MeV \pm 10mm)}/N^{targ}_{e^+}$) & 0.16   \\
Compton source efficiency($N^{\pm10 MeV \pm10mm}_{e^+}/N_{e^-}$) & 0.0051   \\
 \bottomrule
 \end{tabular}
\label{table:results}
\end{table}

\section{Conclusions}
In the context of the polarized positron sources, we simulated the possible layout of Compton polarized positron source using the multiple IP line. 
The main positron losses are occurred after the AMD and first TW structure. So, to improve the CPS efficiency, the further optimizations of the ACS can be done.  One of them can imply the tuning the RF phase of the first TW structure to decelerate more the positrons in the first structure and further accelerate them in the following ones. Such technique can allow to capture more positrons and improve the positron bunch characteristics like the energy spread/bunch length what facilitates the stacking of the positrons bunches into the DR.

\end{document}